\documentclass[a4paper]{PoS}

\title{Bulk viscous corrections to screening and damping in the
  deconfined phase at high temperature}

\ShortTitle{Bulk viscous corrections to screening and damping}

\author{\speaker{Adrian Dumitru}\thanks{I thank CPOD 2017 for the
    opportunity to present this work and Qianqian Du, Yun Guo, and
    Michael Strickland for their collaboration on the material
    presented in secs.~\protect\ref{sec:Pi} and~\protect\ref{sec:V}. I also thank Ho-Ung
    Yee and S.~Mukherjee for useful comments at the meeting. I am
    grateful for support by the
    DOE Office of Nuclear Physics through Grant
    No.\ DE-FG02-09ER41620; and from The City University of New York
    through the PSC-CUNY Research grant 60262-0048.}\\
  Department of Natural Sciences, Baruch College, CUNY, 17 Lexington Avenue, New
  York, NY 10010, USA\\
  The Graduate School and University Center, The
  City University of New York, 365 Fifth Avenue, New York, NY 10016, USA\\
  Physics Department, Brookhaven National Lab, Upton, NY 11973, USA\\
  E-mail: \email{adrian.dumitru@baruch.cuny.edu}}

\abstract{Non-equilibrium corrections in a hot QCD medium modify the
  ``hard thermal loops'' (HTL) which determine the resummed
  propagators for gluons with soft momenta as well as the Debye
  screening and Landau damping mass scales. We focus on bulk viscous
  corrections to a thermal fixed point. The screening and damping mass
  scales are sensitive to the bulk pressure and perhaps to (pseudo-)
  critical dynamical scaling of the bulk viscosity in the vicinity of
  a second-order critical point. This would affect the properties of
  quarkonium bound states in the deconfined phase. }

\FullConference{Critical Point and Onset of Deconfinement - CPOD2017\\
		7-11 August, 2017\\
		The Wang Center, Stony Brook University, Stony Brook, NY}

\def\be{\begin{equation}}
\def\ee{\end{equation}}
\def\bea{\begin{eqnarray}}
\def\eea{\end{eqnarray}}
\def\simle{\mathrel{
    \rlap{\raise 0.511ex \hbox{$<$}}{\lower 0.511ex \hbox{$\sim$}}}}
\def\simge{\mathrel{
    \rlap{\raise 0.511ex \hbox{$>$}}{\lower 0.511ex \hbox{$\sim$}}}}

\begin{document}

\section{Introduction, bulk viscosity in QCD}
The bulk viscosity $\zeta(T)$ in QCD at very high temperatures
$T\gg\Lambda_{\rm QCD}$ has been computed to leading order in the
coupling in Ref.~\cite{Arnold:2006fz}. They find that it is very small
since $\zeta$ is proportional to the {\em square} of the
deviation from conformality given by the $\beta$-function. This leads
to $\zeta/\eta\sim \alpha_s^4$ (neglecting logarithms of the inverse
coupling) which is proportional to
\be \label{eq:alpha4_cs}
(\alpha_s^2N^2)^2\sim (1-3c_s^2)^2~.
\ee
$c_s$ is the speed of sound. To see this write the trace anomaly
in the pure glue theory in the form~\cite{TraceAnomalyFiniteT}
\be \label{eq:TrAnom}
\beta(\alpha_R) \frac{\partial}{\partial\alpha_R} p(T,\alpha_R) = e-3p~.
\ee
Here, $\alpha_R$ denotes the renormalized coupling at a scale $\mu$
and $p(T,\alpha_R)$ is the pressure as a function of $T$ and
$\alpha_R$. On the l.h.s.\ the derivative w.r.t.\ the renormalized
coupling is performed at constant $T$. In (pure glue) perturbation
theory, up to first order in the running coupling
$\alpha_s(T;\alpha_R)$:
\be \label{eq:p_alphas}
p(T,\alpha_R) - p(T,0) = - \frac{N(N^2-1)}{144} \alpha_s(T) \, T^4
\ee
and eq.~(\ref{eq:TrAnom}) turns into
\bea
e-3p &=& - \frac{N(N^2-1)}{144} T^4 \beta(\alpha_R)
\frac{\partial}{\partial\alpha_R} \alpha_s(T)~, \nonumber\\
&=& - \frac{N(N^2-1)}{144} T^4 \beta(\alpha_s(T))~. \label{eq:e-3p_beta}
\eea
Differentiate both sides w.r.t.\ the energy density; on the r.h.s.\ write
this derivative as $(\partial e/\partial T)^{-1}\, \partial/\partial
T$ and note that to leading order this latter derivative w.r.t.\ $T$
acts on the factor $T^4$ only:
\bea
1-3c_s^2 &=&
- \frac{\frac{N(N^2-1)}{36} T^3 \beta(\alpha_s(T))}{\partial e/\partial T}~.
\eea
Thus, $1-3c_s^2$ is indeed proportional to $-N\beta(\alpha_s(T))\sim
\alpha_s^2(T)\, N^2$.

On the other hand, when $(e-3p)/T^4$ exhibits a power law tail then
the proportionality~(\ref{eq:alpha4_cs}) can be modified. For
illustration consider a simple model where a non-perturbative
contribution~\cite{p_nonpt} is added to the one-loop pressure while
the two-loop pressure remains\footnote{The assumption that the
  two-loop pressure is unaffected by the non-perturbative contribution
  does {\em not} correspond to the models proposed in
  refs.~\cite{p_nonpt}.} the one from eq.~(\ref{eq:p_alphas}):
\be \label{eq:p_npt}
p(T) = (N^2-1)\frac{\pi^2}{45}\left( T^4 - T^2 T_c^2\right)
- \frac{N(N^2-1)}{144} \alpha_s(T) \, T^4~.
\ee
At order ${\alpha_s}^0$ this gives
\bea
e-3p &=& 2(N^2-1)\frac{\pi^2}{45} T^2 T_c^2\approx 0.44 (N^2-1) T^2
T_c^2~, \label{eq:e-3p_T2Tc2} \\
c_s^2 &=& \frac{2T^2-T_c^2}{6T^2-T_c^2}~. \label{eq:cs2_T2Tc2}
\eea
$(e-3p)/T^2T_c^2$ in fact agrees quite well with (pure gauge) lattice
data~\cite{Latt_e-3p_N=3,Latt_e-3p_N>3} for $N=3,4,6$ colors in the
regime $T/T_c\simeq 1.5 - 4$. One can now obtain the running coupling
$\alpha_s(T)$ from the trace anomaly. To do so, replace the l.h.s.\ of
eq.~(\ref{eq:e-3p_beta}) by (\ref{eq:e-3p_T2Tc2}):
\be
N\beta(\alpha_s(T)) = - \frac{288\pi^2}{45} \frac{T_c^2}{T^2}
~~~~\rightarrow ~~~~
N\alpha_s(T) = \frac{144\pi^2}{45} \frac{T_c^2}{T^2}~~~~,~~~~
\beta(\alpha_s) = - 2 \alpha_s~.
\ee
This result for the running coupling shows explicitly why this
illustration is not self consistent: in eq.~(\ref{eq:p_npt}), the
contribution to the pressure at order $\alpha_s$ is comparable to the
leading contribution, just like the pure perturbative expansion of the
pressure. Nevertheless, to complete this toy model calculation use
eq.~(\ref{eq:cs2_T2Tc2}) to express $1-3c_s^2$ in terms of
$\alpha_s(T)$:
\be
1-3c_s^2(T) = \frac{90N\alpha_s(T)}{864-45N\alpha_s(T)}~.
\ee
Hence, at $T/T_c=4$, say, where $\alpha_sN$ is sufficiently small,
$1-3c_s^2$ would be approximately linear instead of quadratic in
$\alpha_sN$ (and linear in $-\beta N$).

In a more realistic theory of the deconfined phase one should not
expect that $1-3c_s^2(T)$ is given simply by a number times a power of
$\alpha_s(T)$. Rather, that ``number'' presumably would be a
dimensionless function $f(T/T_c, T/m_D, \dots)$ of
temperature. However, if that theory does not involve or generate
$T_c$ then a fit of the form $1-3c_s^2(T) = \mathrm{const.} \times
(\alpha_s(T) N)^{b(T)}$ would still be interesting. In particular, a
fit to $N_f=2+1$ three loop HTLpt over $T=0.15\to 1.5$~GeV (at
$\mu_B=0$) with constant power $b$ gives $b\approx 3.3$~\cite{HTLpt}.

For completeness we mention also that in the limit of $N\to \infty$ and
large 't~Hooft coupling $\lambda\gg1$ the holographic correspondence
for broken conformal invariance suggests $\zeta/\eta \sim
1-3c_s^2$~\cite{Buchel:2007mf} in 3 spatial dimensions; also see
refs~\cite{HoloBulk}.

For heavy-ion collisions the most relevant temperature regime is
$T\simle 4T_c$.  The lattice has shown that the trace anomaly of
QCD, expressed as energy density minus three times the pressure, grows
large at $T\sim T_c$~\cite{Bazavov:2014pvz}.  Thus, it
has been suggested in the literature that the bulk viscosity to
entropy density ratio should increase, too, as the temperature
approaches the confinement-deconfinement
temperature~\cite{KharzeevTuchin}.  Here we perform a weak-coupling
HTL computation to assess, at least qualitatively, the impact of
bulk-viscous corrections on the heavy-quark potential. This analysis
does not apply for $T\simeq T_c$ but it provides
baseline expectations for bulk-viscous effects on screening and damping
from (resummed) weakly coupled QCD.

Bulk viscous corrections are expected to become important also in the
vicinity of a second order critical point; this could be realized in
hot QCD either by tuning of the quark masses~\cite{qmass} or perhaps
by introducing a baryon charge asymmetry~\cite{muB_critical_point}.
Due to critical slowing down the bulk viscosity should
diverge~\cite{Moore:2008ws} $\zeta\sim \xi^z$ where $\xi\to\infty$ is
the correlation length and $z$ is a dynamical critical
exponent. However, since the relaxation time in the critical region of
the bulk pressure diverges as well, in heavy-ion collisions it
is not expected to exceed the ideal pressure~\cite{Monnai:2016kud}.

\section{HTL resummed gluon self energies incl.\ bulk-viscous
  corrections} \label{sec:Pi}

To compute the gluon self energy in the hard loop approximation we
require the phase space distributions of the particles in the medium.
In the local rest frame, we take them to be
\be
f({\bf p}) = f_{\rm id}(p) + \delta_{\rm bulk} f(p) + \delta_{\rm shear}
f({\bf p})~.
\label{eq:dis}
\ee
Here, $f_{\rm id}(p)$ is an isotropic reference distribution when
non-equilibrium corrections are absent. This would normally correspond
to thermal Fermi-Dirac or Bose-Einstein distributions, respectively,
if the ``ideal'' reference is the thermal fixed point.

The corrections $\delta f$ in Eq.~(\ref{eq:dis}) correspond to non-equilibrium
corrections. We take the isotropic correction $\delta_{\rm bulk}
f(p)$ as a bulk-viscous correction while the anisotropic part
$\delta_{\rm shear} f({\bf p})$ is analogous to shear. However, we do
not assume that these corrections are parametrically suppressed. The
corrections to the real and imaginary parts of the HTL resummed gluon
propagator due to $\delta_{\rm shear} f({\bf p})$ have been worked out
in Refs.~\cite{Dumitru:2007hy}.  Here, we focus on bulk
viscous corrections instead.

To obtain explicit results for the self energies we require a model
for $\delta_{\rm bulk} f(k)$. We assume that the bulk viscous
correction to the local thermal distribution function takes the form
\be
\delta_{\rm bulk}    f(k)=
\left(\frac{k}{T}\right)^a \Phi\, f_{\rm id}(k)(1\pm
f_{\rm id}(k))~. \label{bulk:discorrection}
\ee
$\Phi$ is proportional to the bulk pressure, which in the
Navier-Stokes / Landau approximation is negative, and $a$ is a constant. We
require $a>0$ to ensure that the dominant contribution to the retarded
self energy is from hard (gluon) loop momenta, $k\sim T$.  The bulk
viscous correction to the symmetric self energy at ${\cal{O}}
(\Phi^2)$ involves the fourth power of the distribution function and
requires a more stringent bound, $a>1/2$. It should be clear
that~(\ref{bulk:discorrection}) is a generic (and simple) model for
the non-equilibrium correction chosen such as to maintain
applicability of HTL power counting which may in principle be violated
in certain non-equilibrium scenarios. Furthermore, we recall that if
$\delta_{\rm bulk} f(k)$ does not vanish then the hard scale $T$ in
the distributions functions is no longer equal to the temperature
which instead must be determined from the Landau matching
condition. However, the hard scale remains on the order of the
temperature and we continue to denote it as $T$ for simplicity of
presentation. Finally, we shall further assume that $|\Phi|\gg g^2$ so
that two-loop corrections to the gluon self energy are negligible.

Using standard methods of real-time thermal field theory~\cite{Mrowczynski:2000ed}
one can derive~\cite{Du:2016wdx} the following expressions for the retarded and
symmetric time-ordered (temporal) gluon self energies\footnote{For
  brevity we restrict to vanishing quark chemical potential, see
  ref.~\cite{Du:2016wdx} for the corresponding expressions at $\mu>0$.}:
\begin{eqnarray}
 \Pi_R(P) &=& \frac{N_f g^2}{(2\pi)^2}\int k d k \, f_F(
k) \int d \Omega_k \frac{1-({\hat{\bf k}}\cdot {\hat{\bf p}})^2}{({\hat{\bf
       k}}\cdot {\hat{\bf p}}+\frac{p_0+i\, \epsilon}{p})^2}~, \label{general:re}\\
\Pi _F(P) &=& - i \frac{N_f g^2}{\pi} \frac{2}{p}\Theta(p^2-p_0^2) \int
k^2dk f_F(k)(1-f_F(k)) \, .
\label{general:sy}
\end{eqnarray}
The advanced self energy is equal to $\Pi_R(P)$ with an inverted sign of $\epsilon$.
These expressions account for the contribution due to $N_f$ loops of
massless quarks. For the contribution due to a gluon loop replace
the Fermi-Dirac distribution $f_F(k)$ by a Bose distribution $f_B(k)$,
Pauli blocking $1-f_F(k)$ by Bose enhancement $1+f_B(k)$, and $N_f$ by
$N_c$. Here, $f(k)$ corresponds to the non-equilibrium distribution
function introduced in eqs.~(\ref{eq:dis},\ref{bulk:discorrection}) above.

It is clear from eqs.~(\ref{general:re},\ref{general:sy}) that for
isotropic $\delta f$ the dependence of the self energies on the
external energy/momentum is the same as in equilibrium. However, there
is a correction to the mass scales which appear in $ \Pi_R(P)$ and $ \Pi_F(P)$:
\begin{eqnarray}
m_R^2 = \left(2N_c+N_f\right) \frac{g^2 T^2}{6} &\rightarrow&
m_{R}^2 + \delta m_{R}^2 = \nonumber \\
& & \left(2N_c\left(1+c_R^{(g)}(a)\Phi\right)
+N_f\left(1+c_R^{(q)}(a)\Phi\right)\right)
\frac{g^2 T^2}{6}~, \label{eq:shift_mD_Pi_R} \\
m_F^2 = \left(2N_c+N_f\right)
\frac{g^2 T^2}{6} &\rightarrow&
m_{F}^2 + \delta m_{F}^2 = \nonumber\\
& & \left(2N_c\left(1+c_F^{(g)}(a)\Phi\right)
+N_f\left(1+c_F^{(q)}(a)\Phi\right)\right)
\frac{g^2 T^2}{6}~.\label{eq:shift_mD_Pi_F}
\end{eqnarray}
In~(\ref{eq:shift_mD_Pi_F}) only the correction linear in $\Phi$ has
been given, see ref.~\cite{Du:2016wdx} for $m_{F}^2 + \delta m_{F}^2$
to order $\Phi^2$.  Here,
\begin{eqnarray}
c_R^{(q)}(a) &=& 2 (1-2^{-a}) \, c_R^{(g)}(a) =
\frac{12}{\pi^2}(1-2^{-a})\, \Gamma(2+a)\, \zeta(1+a)~, \\
c_F^{(q)}(a) &=& 2 (1-2^{-a}) \, c_F^{(g)}(a) =
\frac{6}{\pi^2}(1-2^{-a})\, \Gamma(3+a)\, \zeta(1+a)~,
\label{eq:cRq}
\end{eqnarray}
are pure numbers of order 1. Therefore, a substantial (negative)
bulk-viscous pressure $\Phi$ could potentially ``short out'' the
self-energies. It would be interesting to check the values of $m_R^2$
and $m_F^2$ obtained in a hydrodynamic simulation which incorporates
the critical behavior of the bulk viscosity (such as presented in
ref.~\cite{Monnai:2016kud}). Non-equilibrium effects on the gluon self
energies (or the heavy-quark potential) should also be visible in
transport theory approaches~\cite{Rapp}.

The Schwinger-Dyson equation determines the HTL resummed propagators
(in Coulomb gauge),
\begin{eqnarray}
{{D}^*}_{R}(P) &=& \frac{1} {p^2 -
  {\Pi}_{R}(P)} = \frac{1} {p^2 - \left(m_{R,D}^2 +
  \delta m_{R,D}^2\right)\left (\frac{p_0}{2 p}\ln
  \frac{p_0+p+i\epsilon} {p_0-p+i\epsilon}
  -1\right)}~,  \label{probc}\\
{{D}^*}_{F}(P) &=&  {{D}^*}_{R}(P)\,{\Pi} _{F}(P)  \,
{{D}^*}_{A}(P) \, . \label{dft}
\end{eqnarray}
The advanced propagator is equal to ${{D}^*}_{R}(P)$ with a reversed
sign of $\epsilon$.

\section{Non-equilibrium corrections to the HTL static potential}
\label{sec:V}

We can compute the static potential due to one gluon exchange through
the Fourier transform of the physical ``11" Schwinger-Keldysh
component of the (longitudinal) gluon propagator in the static limit,
\begin{eqnarray}
V({\bf{r}}) &=& (ig)^2 C_F\int \frac{d^3{\bf{p}}}{(2\pi)^3} \,
\left(e^{i{\bf{p \cdot r}}}-1\right)\, \left({{D}^*}(p_0=0,
  \bf{p})\right)_{11} \nonumber\\
&=& -g^2 C_F\int \frac{d^3{\bf{p}}}{(2\pi)^3} \, \left(e^{i{\bf{p \cdot
r}}}-1\right)\,
\frac{1}{2}\left({{D}^*}_R+{{D}^*}_A+{{D}^*}_F\right)~. \label{eq:Vr}
\end{eqnarray}
We have also subtracted an $r$-independent (but $T$-dependent)
self-energy contribution.  The Fourier transform of the sum of
retarded and advanced propagators gives a real Debye screened
potential,
\be \label{eq:Re_V}
\mathrm{Re}\, V(r) = - \frac{g^2 C_F}{4\pi r}\, e^{-\hat{r}} ~,
\ee
where $\hat{r} \equiv r \sqrt{m_{R}^2 + \delta m_{R}^2}$.
The imaginary part of the potential originates from the symmetric
propagator ${{D}^*}_F$ and is due to Landau damping of the gluon
exchanged by the static sources~\cite{Laine:2006ns},
\begin{equation}
\mathrm{Im}\, V(r) = - \frac{ g^2 C_F T}{4 \pi } \frac{m_{F}^2 +
  \delta m_{F}^2}{m_{R}^2 + \delta m_{R}^2} \,
\phi(\hat{r})\, , \label{eq:ImVr}
\end{equation}
with $\phi(\hat{r})\sim \hat{r}^2\,{\log}\, 1/\hat{r}$ when
$\hat{r}\ll1$. The imaginary part of the potential generates a thermal
width for quarkonium bound states.

Hence, bulk-viscous corrections in general modify the Debye screening
of the $1/r$ Coulomb potential. Also, $\mathrm{Im}\, V(r)$ is
multiplied by $(m_{F}^2 + \delta m_{F}^2)/(m_{R}^2 + \delta
m_{R}^2)$ which we expect to be less than 1 typically. Thus, a
significant bulk pressure should reduce the thermal width of
quarkonium states. The potential written in
eqs.~(\ref{eq:Re_V},\ref{eq:ImVr}) is not expected to provide accurate,
quantitative predictions of the properties of quarkonium bound states
at finite temperature. However, basic qualitative insight obtained
with the methods described here can be useful for incorporating
non-equilibrium effects into quarkonium potentials derived from
lattice QCD~\cite{RthkpfStklnd}.

\end{document}